\documentclass[prb,twocolumn,showpacs,preprintnumbers,amsmath,amssymb]{revtex4}
\usepackage{graphicx}
\usepackage{longtable}
\usepackage{dcolumn}
\usepackage{bm}
\usepackage{color}

\begin{document}

\title{Itinerant G-type antiferromagnetism in D0$_3$-type V$_3$Z (Z=Al, Ga, In) compounds: \\
A first-principles study}

\author{Iosif Galanakis$^{1}$}\email{galanakis@upatras.gr}
\author{\c{S}aban T{\i}rpanc{\i}$^{2}$}
\author{Kemal \"{O}zdo\u{g}an$^{3}$}
\author{Ersoy \c{S}a\c{s}{\i}o\u{g}lu$^{4}$}\email{e.sasioglu@gmail.com}

\affiliation{$^{1}$Department of Materials Science, School of Natural Sciences,
University of Patras,  GR-26504 Patra, Greece \\
$^{2}$Department of Physics, Gebze Technical University
41400 Gebze, Kocaeli, Turkey \\
$^{3}$Department of Physics, Yildiz Technical University,
34210 \.{I}stanbul, Turkey \\
$^{4}$Peter Gr\"{u}nberg Institut and Institute for
Advanced Simulation, Forschungszentrum J\"{u}lich and JARA, D-52425
J\"{u}lich, Germany}

\date{\today}

\begin{abstract}
Heusler compounds are widely studied due to their variety of
magnetic properties  making them ideal candidates for spintronic
and magnetoelectronic applications. V$_3$Al in its metastable
D0$_3$-type Heusler structure is a prototype for a rare
antiferromagnetic gapless behavior. We provide an extensive study
on the electronic and magnetic properties of V$_3$Al, V$_3$Ga and
V$_3$In compounds based on state-of-the-art electronic structure
calculations. We show that the ground state for all three is a
G-type itinerant antiferromagnetic gapless semiconductor. The
large antiferromagnetic exchange interactions lead to very high
N\'eel temperatures, which are predicted to be around 1000 K. The
coexistence of the gapless and antiferromagnetic behaviors in
these compounds can be explained considering the simultaneous
presence of three V atoms at the unit cell using arguments which
have been employed for usual inverse Heusler compounds. We expect
that our study on these compounds to enhance further the interest
on them towards the optimization of their growth conditions and
their eventual incorporation in devices.
\end{abstract}


\maketitle

\section{Introduction}
\label{sec1}

The constant growth of computational materials science triggered
an even more exciting growth in experimental materials science.
Several phenomena have been explained based on ab-initio
electronic structure calculations and several compounds with
predefined properties targeting at specific applications have been
studied. Among them exist the so-called half-metallic
Heusler \cite{Heusler} compounds a special class of magnets which
present semiconducting behavior for one of the two spin
channels.\cite{FelserRev,FelserRev2} Several Heusler compounds
have been identified using ab-initio calculations prior to their
experimental growth.\cite{Gillessen2008,Gillessen2009} Such
materials can find a variety of applications in the field of
magnetoelectronics (nanoelectronics where only magnetic materials
are employed) and spintronics (nanoelectronics where hybrid
devices of semiconductors and magnetic materials are
used).\cite{ReviewSpin} We should note here that there are also
other known half-metals like La$_{0.7}$Sr$_{0.3}$MnO$_3$ which
have been also demonstrated to present fully spin-polarized
tunnelling in magnetic tunnel junctions,\cite{Bowen} but Heusler
compounds remain very attractive for applications due to their
very high Curie temperatures and their structural similarity to
binary semiconductors.\cite{landolt,landolt2}

The most studied Heusler compounds in literature are the so-called
full Heuslers having the chemical formula X$_2$YZ, like
Co$_2$MnSi, and several have been identified as
half-metals.\cite{Galanakis2002b} When the valence of the X is
smaller than the valence of the Y they are called inverse Heusler
compounds and crystallize in a similar structure where only the
sequence of the atoms changes.\cite{GSP} Also most of the latter
compounds are half-metallic magnets.\cite{GSP} Although the
compounds referred to above can find several applications in
spintronics and magnetoelectronics,\cite{Perspectives} several
half-metallic Heusler compounds with more exotic properties have
been found which can further optimize the operation of devices.
Among them are the so-called spin-gapless semiconducting
(SGS)\cite{Wang} Heuslers which present a usual semiconducting
band structure for one spin direction and a gapless (almost or
exactly zero energy gap) in the other
spin-direction.\cite{GalanakisSGS,Jakobsson} Such materials can
enhance the performance of devices since vanishing energy is
needed to excite both electrons and holes and a prototype
Mn$_2$CoAl has been already grown experimentally and its
SGS properties have been confirmed.\cite{Ouardi}

\begin{figure}[t]
\includegraphics[width=\columnwidth]{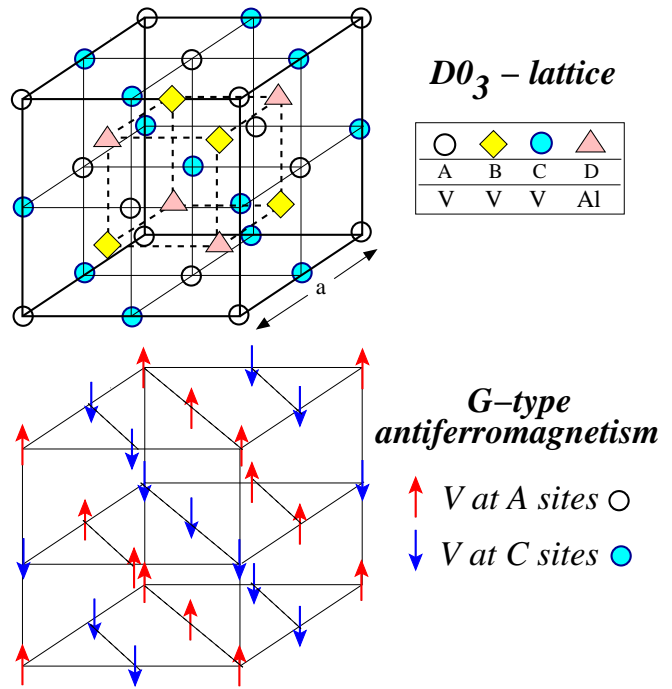}
 \caption{(Color online) Upper panel: Schematic representation of
the cubic D0$_3$ structure adopted by the V$_3$Z Heusler
compounds. The unit cell is that of an f.c.c. with four
equidistant atoms as basis along the [111] diagonal. The atoms at
the A and C sites are at the center of a cube surrounded by four
atoms at the B and four atoms at the C sites and vice versa. The V
next-nearest neighboring atoms at the A and C sites for a lattice
of octahedral symmetry if we neglect the B and D sites.\\
Lower panel: Schematic representation of the G-type
antiferromagnetic order made up of successive (111) planes which
are antiferromagnetically coupled. The V atoms at the A and C
sites have antiparallel spin magnetic moments denoted by arrows of
opposite direction. The V atoms at the B sites and the Al atoms
have zero spin magnetic moments due to symmetry reasons being
situated at the center of cubes having four V atoms at the A sites
and four V atoms at C sites as nearest neighbors.} \label{fig1}
\end{figure}

A material of special interest is V$_3$Al, a Heusler compound
crystallizing in the so-called D0$_3$ lattice structure shown in
Fig.\,\ref{fig1} which resembles the cubic structure of
full-Heuslers where now all X an Y atoms are identical adopted
also by other materials like Fe$_3$Al, Fe$_3$Si, and
Cr$_3$Se.\cite{Fe3AlSi,Cr3Se} The first attempt to study this
material using first-principles calculations predicted a
non-magnetic ground state.\cite{Gao13} But latter calculations by
Skaftouros and collaborators predicted that the ground state is in
reality an antiferromagnetic gapless
semiconductor\cite{GalanakisSGS} resembling the well-known gapless
semiconductors.\cite{GS} Such an electronic structure is possible
since as shown in Fig.\,\ref{fig1} there are two V atoms sitting
at the A and C sites which form a simple cubic structure, if we
neglect the other sites. The V atoms at these sites are allowed
due to symmetry to have antiparallel spin magnetic moments of the
same size leading to G-type antiferromagnetism shown schematically
also in Fig.\,\ref{fig1}. The V atoms at the B sites and the Al
atoms at the D sites are at the center of a cube surrounded by
four V atoms at A sites and four V atoms at C sites and thus due
to symmetry reasons their spin magnetic moment in such a
configuration should be zero. This is compatible with the
so-called Slater-Pauling rule connecting the total number of
valence electrons to the total spin magnetic moment in Heusler
compounds.\cite{GSP} In 2015 Jamer and collaborators presented an
extensive study on V$_3$Al combining both electronic structure
calculations and experiments.\cite{Jamer} On one hand their
simulations confirmed the results of Skaftouros \emph{et al}, and
on the other hand they have successfully grown films of V$_3$Al
and dichroic experiments using synchrotron radiation were
compatible with an antiferromagnetic state. Here we have to note
that the ground state of V$_3$Al is not the Heusler structure but
the A15 lattice structure and V$_3$Al in this structure is a
well-known superconductor.\cite{A15,A15b}

V$_3$Al can be viewed as a prototype material for studying gapless
antiferromagnetic behavior and can be considered as a cornerstone
for future spintronic and magnetoelectronic devices based on
antiferromagnetic elements. Thus in  present paper we provide an
extensive study based on simulations of the electronic and
magnetic properties of V$_3$Al as well as of the stability of its
antiferromagnetic character. To make our study more complete we
have also included results on V$_3$Ga and V$_3$In compounds which
have the same number of valence electrons. Also these compounds
were found to be antiferromagnetic gapless semiconductors and thus
we will mainly concentrate on V$_3$Al but conclusions are also
valid for them. In Sec.\,\ref{sec2} we shortly present the
computational method. In Sec.\,\ref{sec3-1} we discuss the
electronic properties and the gapless behavior of the V$_3$Al
compound under study and in Sec. \ref{sec3-2} its magnetic
properties including also the calculation of the exchange
constants and the N\'eel temperature. Sec.\,\ref{sec3-3} is
devoted to the origin of the gapless behavior. Finally in
Sec.\,\ref{sec4} we summarize and present our conclusions. We
believe that our present results will even further intensify the
interest on this unique compound.

\section{Computational method}
\label{sec2}

To perform the electronic structure calculations, we employed the
full-potential nonorthogonal local-orbital minimum-basis band
structure scheme (FPLO)\cite{FPLO,FPLO2} within the generalized
gradient approximation (GGA) as parameterized by Perdew, Burke and
Ernzerhof.\cite{GGA} Some of the results presented in
Secs.\,\ref{sec3-1} and \ref{sec3-2} have been also obtained using
the local-spin-density approximation (LSDA).\cite{LSDA} We have
used the lattice parameter of 6.09 \AA\  calculated via total
energy calculations also using the FPLO method within GGA in
Ref.\,\onlinecite{GalanakisSGS}. Using the same method we have
also calculated the equilibrium lattice constants for the other
two compounds and found a value of 6.07 \AA\ and 6.32 \AA\ for
V$_3$Ga and V$_3$In, respectively (see Table\,\ref{table}).  For
the integrations in the first Brillouin zone a dense
Monkhorst-Pack grid has been used.\cite{Monkhorst} All results
presented in this study have been obtained using FPLO with the
exception of the exchange constants and N\'eel temperature in
Sec.\,\ref{sec3-2}, which are obtained employing the ASW
method.\cite{asw}.

To calculate interatomic  exchange parameters we employ the
frozen-magnon technique\cite{magnon1,magnon2,magnon3} as
described in Refs.\,\onlinecite{Sasioglou2004}  and \onlinecite{SasiogluH1}.
The N\'eel temperature is estimated by employing the so-called
random-phase-approximation (RPA) approach.\cite{tyablikov,Callen,SasiogluH1,pajda,bouzerar}

\section{Results and Discussion}
\label{sec3}

\begin{figure}[t]
\includegraphics[width=\columnwidth]{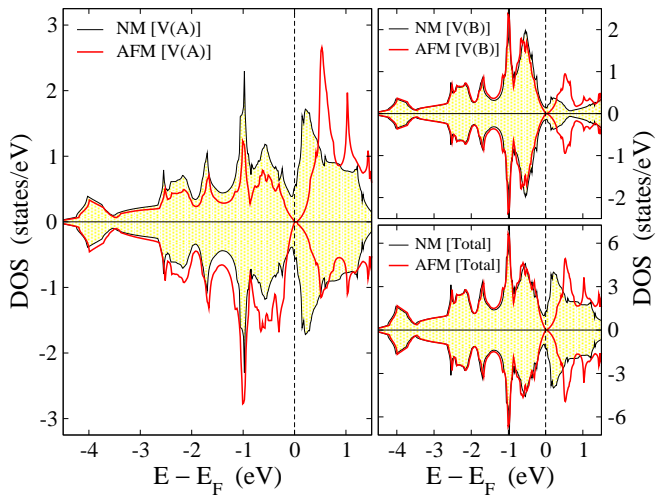}
\vspace{-0.2 cm}
\caption{(Color online)  Total and atom-resolved density of states (DOS) within GGA
for the non-magnetic (NM) and antiferromagnetic (AFM) magnetic
configurations of $V_3$Al. Positive (Negative) DOS values
corresponds to the spin-up (spin-down) electrons. The Fermi level
corresponds to the zero energy.} \label{fig2}
\end{figure}

\subsection{Electronic properties and gapless behavior} \label{sec3-1}

\begin{figure}[t]
\includegraphics[width=\columnwidth]{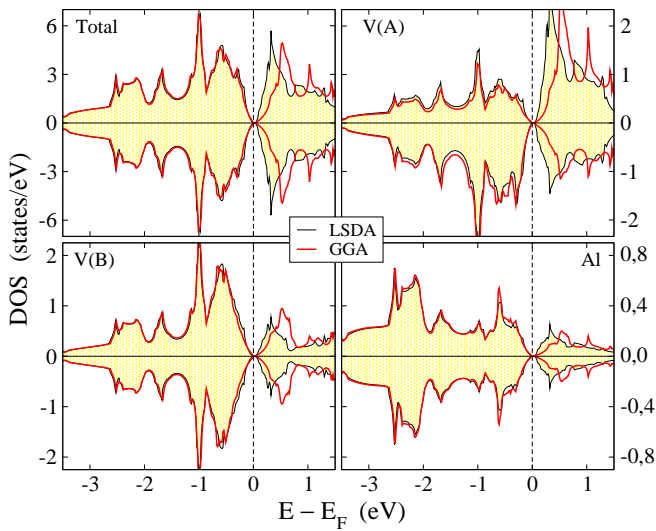}
\vspace{-0.2 cm} \caption{(Color online)  GGA vs LSDA calculated
DOS for the antiferromagnetic V$_2$Al compound. Details as in Fig.\,\ref{fig2}.} \label{fig3}
\end{figure}

First, we should establish the gapless behavior of V$_3$Al. To this
end in Fig.\,\ref{fig2} we have plotted the density of states (DOS)
projected on the V atoms at the A and B sites (the site is written
either in parenthesis or as superscript to distinguish the V
atoms) as well as the total one in the unit cell for both the non
magnetic (NM) and the antiferromagnetic (AFM) configurations. We
should first note that we cannot use the terms majority and
minority spin since there is equal number of electrons of
different spin in the formula unit; we will use the term spin-up
for the positive DOS values and spin-down for the negative DOS
values. We have also chosen the spin-up in the case of the V$^A$
atoms such that its spin magnetic moment in Table\,\ref{table} is
negative. In the AFM case the DOS of the V$^C$ atoms is identical
to the V$^A$ atoms exchanging the spin-up and spin-down electronic
states. In the AFM case one gets a gapless behavior and the
valence and conduction bands touch each other at the Fermi level.
If we compare the NM and AFM calculations, most of the changes
occur at the V$^A$ DOS where the weight of the states around the
Fermi level increases and we have a normal semiconductor. Deeper
in energy the NM and AFM DOS are almost identical. Especially for
the V$^B$ atoms the DOS below the Fermi level is identical for
both NM and AFM calculations. This also may explain the stability
of the AFM case. A close examination of the total DOS reveals that
the NM and AFM DOS are similar throughout most the energy range
but at the Fermi level the zero DOS at the AFM case leads to
smaller values of the total energy stabilizing it against the NM
case where more electronic charge is present at the Fermi level.

\begin{table*}
\caption{Calculated lattice parameters, total energy differences
between non-magnetic and antiferromagnetic states, atom-resolved
spin magnetic moments (in $\mu_{\mathrm{B}}$) and N\'eel
temperatures for D0$_3$-type V$_3$Z (Z=Al, Ga, In) compounds. In
parenthesis we show the LSDA results.}
\begin{ruledtabular}
\begin{tabular}{llllllll}
Compound  &    a(\AA) & $\Delta E$  &
$m_{\texttt{[V]}}^{\texttt{A}}$  &
 $m_{\texttt{[V]}}^{\texttt{B}}$ & $m_{\texttt{[V]}}^{\texttt{C}}$  &
  $m_{\texttt{[Z]}}^{\texttt{D}}$ & $T_\mathrm{N}^{\mathrm{RPA}}$ (K)   \\
\hline
V$_{3}$Al   & 6.09 & -0.12 & -1.65 (-1.12) & 0.00 (0.00) & 1.65 (1.12)  &  0.00 (0.00) & 988 (648)  \\
V$_{3}$Ga   & 6.07 & -0.10 & -1.55 (-1.02) & 0.00 (0.00) & 1.55 (1.02)  &  0.00 (0.00) & 858 (512)  \\
V$_{3}$In   & 6.32 & -0.18 & -1.98 (-1.48) & 0.00 (0.00) & 1.98
(1.48)  &  0.00 (0.00) & 1023 (704)
\end{tabular}
\label{table}
\end{ruledtabular}
\end{table*}

\begin{figure*}[t]
\includegraphics[width=\textwidth]{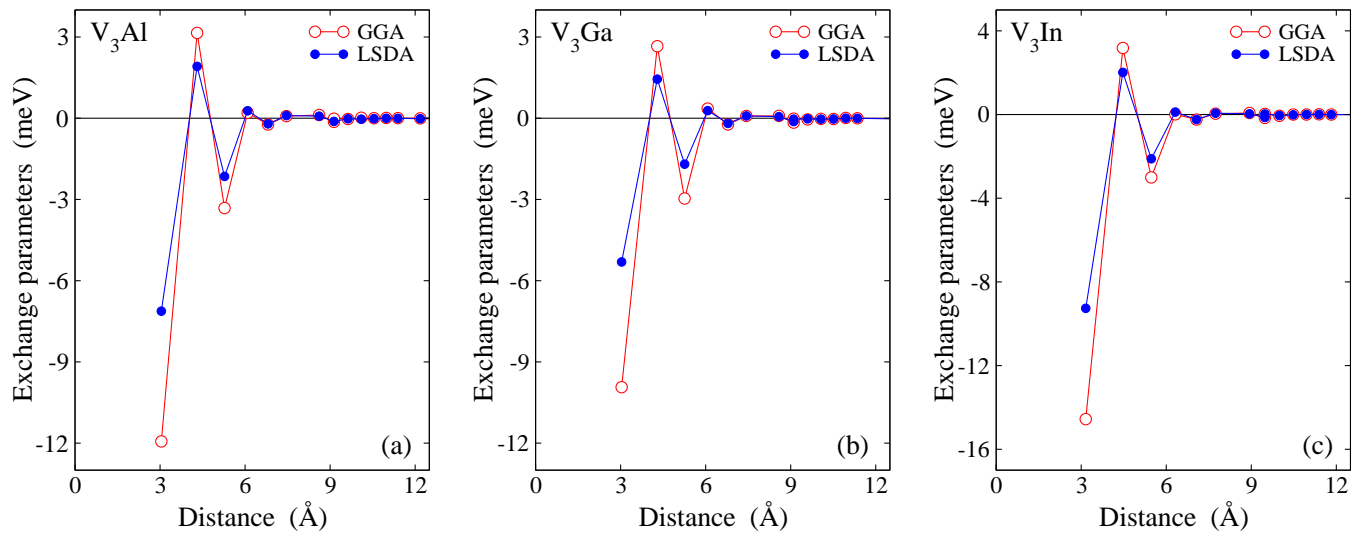}
\caption{(Color online) Interatomic exchange parameters
(inter-sublattice V$^A$-V$^C$ and intra-sublattice V$^A$-V$^A$) as
a function of the distance for all three compounds under study
using both GGA and LSDA functionals.} \label{fig4}
\end{figure*}

To establish the gapless semiconductor we have also performed
calculations using the LSDA functional and present them in Fig.\,\ref{fig3}
versus the GGA results. LSDA is well-known to
underestimate the equilibrium lattice constants and to
overestimate hybridization between orbitals with respect to GGA
but for the same lattice constant both LSDA and GGA should produce
similar electronic properties. This is true since also LSDA
reproduces the gapless semiconducting behavior of GGA. The only
noticeable difference between the two functional is the
distribution of the weight of the unoccupied states just above the
Fermi level. LSDA yields a smaller exchange splitting between
\emph{d}-states of different spin a the V$^A$ and V$^C$ states and
thus there is a small shift of the unoccupied states towards
smaller energy values which also reflects on the DOS of the V$^B$
and Al atoms.

\subsection{G-type antiferromagnetism: stability, magnetic moments,
exchange constants and $T_\mathrm{N}$} \label{sec3-2}

In this subsection we will discuss the magnetic properties of the
compounds under study. In Table\,\ref{table} we have included the
energy difference $\Delta E$ between the non-magnetic and the
antiferromagnetic configurations. For all three compounds $\Delta
E$ is negative  meaning that the AFM state is the ground one
reflecting the discussion in the last paragraph of the previous
section. The values vary between -0.10 and -0.18 eV which are
sizeable and suggest that the magnetic state should be feasible to
stabilize in experiments like the ones of Jamer and
collaborators.\cite{Jamer} Also in Table \ref{table} we have
included the atomic spin magnetic moments. The V$^B$ and Al atoms
have zero spin magnetic moments as expected since they are at the
midpoints between the V$^A$ and V$^C$ atoms. The latter one show
considerable values of atomic spin magnetic moments, whose
absolute values range from 1.55 $\mu_\mathrm{B}$ in the case of
V$_3$Ga to 1.98 $\mu_\mathrm{B}$ for V$_3$In giving a first hint
that exchange interaction should be strong leading to large values
of the N\'eel temperature. Finally, if one looks at the structure
presented in Fig.\,\ref{fig1} neglecting the V$^B$ and Al atoms,
one can consider the structure of being build up by successive
(111) plane made up of either pure V$^A$ or V$^C$ atoms. Thus two
successive (111) planes have atoms of antiparallel spin magnetic
moments and antiferromagnetism is of the so-called G-type shown in
the lower panel of Fig.\,\ref{fig1}.

\begin{figure}[t]
\includegraphics[width=\columnwidth]{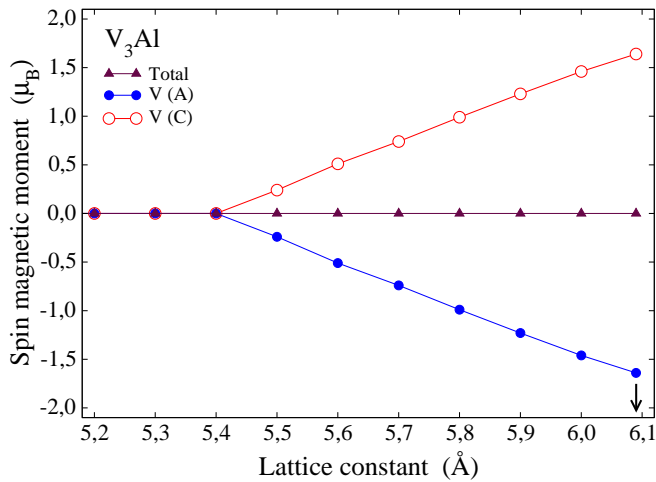}
\vspace{-0.1 cm}
\caption{(Color online) Atomic and total spin magnetic moments in
the case of V$_3$Al upon compression starting from the equilibrium
lattice constant marked by an arrow. Note that spin magnetic moment of the V atom at B site and
Al atom at D site are zero due to symmetry reasons.} \label{fig5}
\end{figure}

In Fig.\,\ref{fig4} we present the calculated interatomic
(intra-sublattice and inter-sublattice) exchange parameters
involving the magnetic V atoms as a function of the distance. We
can easily deduce from the figure that only interactions within
the first three coordination shells are sizable. The first
coordination shell concerns the inter-sublattice V$^A$-V$^C$
exchange interactions (each V$^A$ atom has V$^B$ and Al atoms,
which have zero spin magnetic moment, as nearest neighbors and six
V$^C$ atoms as next nearest neighbors) and it is the largest one
providing the dominating contribution to the formation of the
antiferromagnetic ground state and to the N\'eel temperature. The
second coordination shell refers to the intra-sublattice
V$^A$-V$^A$ exchange interactions and it is of ferromagnetic
character stabilizing further the antiferromagnetic state. Its
value is considerably smaller than the absolute value of the
 exchange interaction between atoms in the 1st coordination
shell due to the large distance between the V$^A$ atoms in the
lattice with respect to the V$^A$-V$^C$ distance. Exchange
interactions quickly decay with distance, which can be attributed
to the existence of the gap in large part of the Brillouin zone
shown in Fig.\,\ref{fig8}. This type of short-range behavior is
typical of most half metallic magnets.\cite{SasiogluH1,Rusz}

RPA estimated N\'eel temperatures,
$T_{\mathrm{N}}^{\mathrm{RPA}}$, within GGA are presented Table
\ref{table}. V$_3$Al shows a $T_{\mathrm{N}}^{\mathrm{RPA}}$ value
of 988 K, V$_3$Ga of 858 K and V$_3$In of 1023 K. All these values
are much larger than the room temperature ensuring that devices
based on these compounds would be functional at room temperature.
The experiments by Jamer and collaborators have provided a value
of about 600 K for V$_3$Al,\cite{Jamer} which is considerably
smaller than our value although it is still high compared to the
room temperature. The discrepancy should be attributed to the
character of the sample in Ref.\,\onlinecite{Jamer}. Our values
concern bulk perfect crystals while experiments have been
performed on polycrystalline films prepared by arc melting. Even
perfect thin films of a material present critical temperatures
much smaller than bulk crystals of the same material and the
discrepancy is much larger for polycrystalline films. Thus in
reality the experimental value is consistent with our theoretical
prediction.

Although the combination of GGA and RPA yields accurate critical
temperature values in Heusler compounds,\cite{SasiogluH1} it would
be interesting to examine also the results of LSDA. As mentioned
above LSDA overestimates the hybridization effect with respect to
GGA resulting in considerable smaller values of the atomic spin
magnetic moments in Table\,\ref{table}. For example the absolute
values of the V atomic spin magnetic moments in V$_3$Al decreases
from 1.65 $\mu_{\mathrm{B}}$ within GGA to 1.12 $\mu_{\mathrm{B}}$
within LSDA a reduction of about 32 \%. The smaller atomic spin
magnetic moments also affect the exchange interactions in
Fig.\,\ref{fig4} which are smaller within LSDA resulting in
smaller predicted values of the $T_{\mathrm{N}}^{\mathrm{RPA}}$ in
Table\,\ref{table}. The discrepancy of the LSDA results with
respect to the GGA ones provides a strong argument towards the
character of the magnetism in the compounds under study. If
magnetism was localized then the hybridization between the
orbitals sitting at nearest sites would be negligible and both
LSDA and GGA should give similar results. When magnetism is of
itinerant character, the hybridization between orbitals of nearest
sites is important and LSDA and GGA would give a sizeable
discrepancy between the computed magnetic properties. This is the
case and thus we can conclude that in V$_3$Al, V$_3$Ga and V$_3$In
compounds the antiferromagnetism is of itinerant character.

\begin{figure}[t]
\includegraphics[angle=-90,width=\columnwidth]{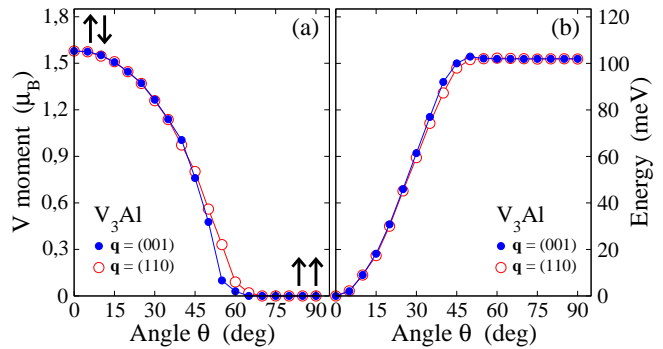}
\caption{(Color online) Behavior of the absolute values of the
atomic spin magnetic moments of the V atoms at the A and C sites
(left panel) and of the total energy (right panel) as a function
of the polar angle $\theta$ (see Ref.\,\onlinecite{Conduction}) for two different values of the
wave vector \textbf{q} (in units of $\frac{2\pi}{\alpha}$) in the
case of the V$_3$Al compound. Notice that by varying $\theta$
between 0$^o$ and 90$^o$ for both wave vectors (001) and (110),
the magnetic structure transforms continuously from the
antiferromagnetic to the ferromagnetic configuration.}
\label{fig6}
\end{figure}

Another characteristic supporting this conclusion is the behavior
of the atomic spin magnetic moments under compression. If their
values persist then magnetism is of localized character, while if
it goes fast to zero magnetism is of itinerant character since
compression fast reduces and eventually kills the hybridization
effect. This has been clearly demonstrated in the case of Cr$_3$Se
where both kind of magnetic behaviors coexist.\cite{Cr3Se} We have
plotted the behavior of the GGA atomic spin magnetic moments under
compression in Fig.\,\ref{fig5}. As we compress the lattice and the
lattice constant is reduced, the absolute values of the spin
magnetic moments of both V$^A$ and V$^C$ atoms present a linear
reduction vanishing at about 5.4 \AA\ and the compound remains a
perfect antiferromagnet under this compression. This behavior is
compatible with the itinerant character of magnetism discussed in
the above paragraph.

Our final step in the investigation of the magnetic properties of
the compounds under study is the stability of the AFM state with
respect to ferromagnetic ordering. In Fig.\,\ref{fig6} we have
plotted for V$_3$Al the variation of the absolute values of the
atomic spin magnetic moments of the V atoms at the A and C sites
(left panel) as a function of the polar angle $\theta$. On the
right panel we plotted the behavior of the total energy as a
function of the polar angle. We took into account two different
values of the wave vector \textbf{q} (in units of
$\frac{2\pi}{\alpha}$): the (001) and (110). The reason is that
for these two values the magnetic structure transforms
continuously from the antiferromagnetic to the ferromagnetic
configuration as the $\theta$ angle changes from 0$^o$ to 90$^o$.
The curves for both \textbf{q} values are similar leading to the
same conclusions. As the angle $\theta$ increases and the spin
magnetic moments rotate to the ferromagnetic configuration the
absolute values of the spin moments decrease and simultaneously
the total energy increases reaching a maximum at about 45$^o$. The
minimum of the total energy corresponds to an angle of 0$^o$ and
thus to the antiferromagnetic coupling of the spin magnetic
moments of the V$^A$ and V$^C$ atoms.

\begin{figure}[t]
\includegraphics[width=\columnwidth]{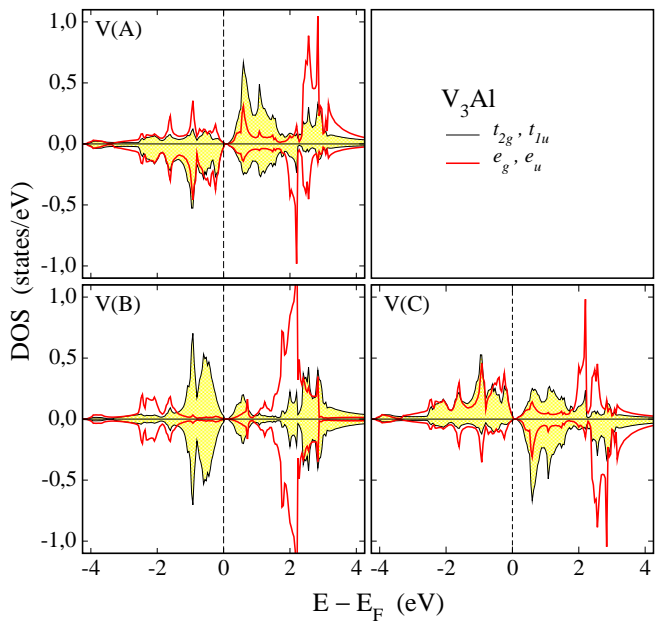}
\vskip -0.1cm \caption{(Color online) DOS for the V atoms in
V$_3$Al projected on the double degenerate $e_\mathrm{g}$ and
$e_\mathrm{u}$, and the triple degenerate $t_\mathrm{2g}$ and
$t_\mathrm{1u}$ d-states. Note that we cannot distinguish between
the $g$-type states obeying both the octahedral and tetrahedral
symmetry and the $u$-type states obeying exclusively the
octahedral symmetry.} \label{fig7}
\end{figure}

\subsection{Origin of the gapless behavior} \label{sec3-3}

In the last part of the discussion of the results, we will
concentrate on the origin of the gapless behavior using arguments
similar to the one used for the full and the inverse Heusler
compounds.\cite{Galanakis2002b,GSP} Again we will use V$_3$Al as
the prototype but results are similar also for the V$_3$Ga and
V$_3$In compounds. First as for the full Heuslers, we can have two
type of \emph{d}-orbitals. The ones concentrated exclusively at
the V$^A$ and V$^C$ sites. These atoms create a simple cubic
structure, if we neglect the V$^B$ and Al atoms, and thus due to
symmetry reasons \emph{d}-hybrids of $u$ type obeying the
octahedral symmetry and being localized exclusively at the V$^A$
and V$^C$ sites are allowed. These can be distinguished between
the double-degenerate $e_u$ and the triple-degenerate $t_{1u}$
states. Except the states of the \emph{u} character we can also
have states of \emph{g} character which obey both the octahedral
and tetrahedral symmetry and are delocalized to all sites. They
also break down to the double-degenerate $e_g$ and the
triple-degenerate $t_{2g}$ states. In the tetrahedral symmetry the
$e_g$ states are lower in energy than the $t_{2g}$ states, while
in the octahedral symmetry it is vice versa and the $t_{1u}$
states are below the $e_u$ states. In Fig.\,\ref{fig7} we have
projected the atom-resolved DOS on the different $g$ and $u$
states. Note that we cannot distinguish the $e_u$($t_{1u}$) from
the $e_g$($t_{2g}$) states. The V$^A$ and V$^C$ atoms posses
states of both character from both sides of the gap. In the case
of V$^B$ atoms, the states deeper in energy are of $e_g$ character
while below the Fermi level states are exclusively of $t_{2g}$
character (note that the $u$ states are not allowed at the V$^B$
site). Above the Fermi level we can find states of both
characters.

\begin{figure}
\includegraphics[scale=0.32,angle=270]{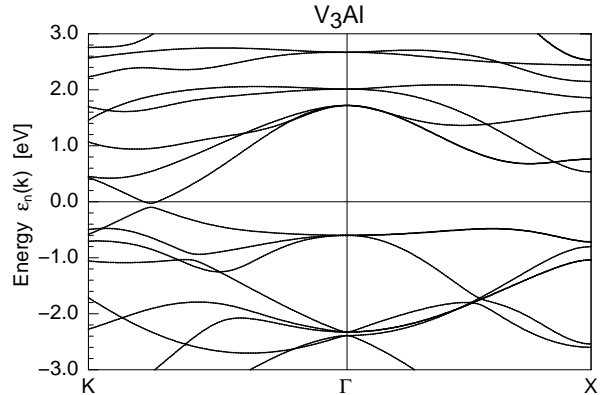}
\vspace{-0.9 cm}
\caption{Band structure along two high-symmetry axis. There is a
direct gap at about one fourth the distance between the K and the
$\Gamma$ high symmetry points. For the character of the bands see
Fig.\,\ref{fig9} and discussion in the text}\label{fig8}
\end{figure}

To make a step further in our understanding of the origin of the
gapless semiconducting behavior we have plotted in Fig.\,\ref{fig8}
the band structure along the K-$\Gamma$ and $\Gamma$-X lines. Note
that the band structure is identical for both spin-directions.
Examining the band structure along several directions (not shown
here) reveals that there is a direct gap of vanishing width
located at about the $\frac{1}{4}$ the K-$\Gamma$ distance. Just
below the Fermi level there  is a triple-degenerate (at the
$\Gamma$-point) band. Deeper in energy at about -2.5 eV below the
Fermi level, at the $\Gamma$-point there are very close in energy
a triple and a double degenerate bands. Lower in energy (not shown
here) is a single band. Above the Fermi level we have a
triple-degenerate (at the $\Gamma$-point) band followed by two
double-degenerate (at the $\Gamma$-point) bands. The character of
the bands at the $\Gamma$-point is primordial for our
understanding of the origin of the gapless behavior since it
reveals also the character of the bands  at the real space and it
has been extensively used in the case of Heusler
compounds.\cite{Galanakis2002b,GSP}

\begin{figure}[t]
\includegraphics[width=\columnwidth]{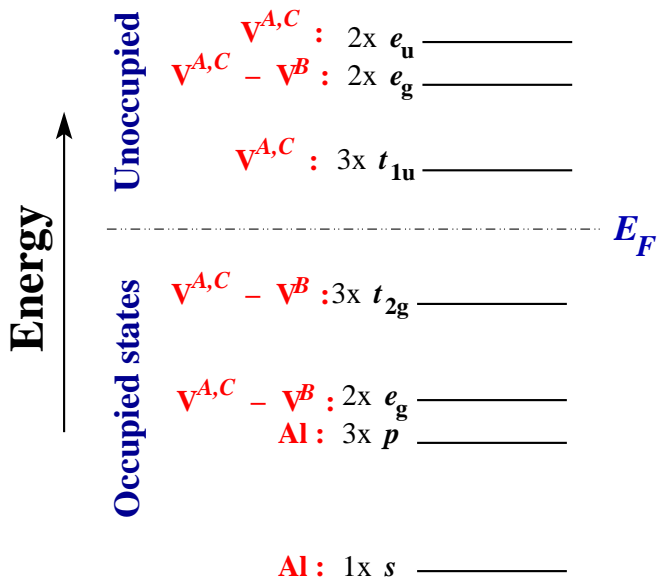}
 \caption{(Color online) Schematic representation of
the character of the bands at the $\Gamma$-point which corresponds
to the character of the orbitals in real space, based on the atom
and orbital resolved bands using the fat band scheme (not shown
here). Notice that the low-lying Al \emph{p} bands accommodate
also a large portion of V's \emph{d} charge and thus have a strong
admixture of the V$^{A,C}$ triple-degenerate $t_{\mathrm{2g}}$
states.} \label{fig9}
\end{figure}

To reveal the character of each band we have performed an analysis
based on the fat band scheme which we have also employed in the
case of the Cr$_3$Se compounds in Ref.\,\onlinecite{Cr3Se}. We do
not show all band structure here but we resume our results in Fig.\,\ref{fig9}
where for one spin-direction we present the character
of the bands. First, we have to note that V$_3$Al has 18 valence
electrons per formula unit (it coincides with the per unit cell
value) and thus per spin we should have 9 occupied states. As for
the minority-band structures of full and inverse Heusler compounds
which show the semiconducting behavior (see Refs.\,\onlinecite{Galanakis2002b} and \onlinecite{GSP}), the single band
low in energy stems from the Al valence \emph{s}-state. At about
-2.5 eV there are very close in energy at the $\Gamma$ point the
triple-band stemming from the Al \emph{p}-states accommodating
also \emph{d} charge from the V atoms, and the double-band
stemming from the $e_g$ states which are spread over all V atoms.
Just below the Fermi level the triple degenerate valence band has
a $t_{2g}$ character and the corresponding orbitals are located at
all V sites; these bands have also a strong Al \emph{p}-admixture
since the $t_{2g}$ states when expressed around the Al site have a
\emph{p} character. The first bands above the Fermi level are the
triple-degenerate $t_{1u}$ bands located mainly at the V$^A$ and
V$^C$ sites followed by the $e_g$ and $e_u$ bands. Thus the Fermi
level is located between the occupied $t_{2g}$  and the empty
$t_{1u}$ states. This resembles the hybridization scheme in the
case of the Sc- and Ti-based inverse Heusler compounds in Ref.
\onlinecite{GSP}. The gap in the latter case is small since it
does not result from a bonding-antibonding hybrid formation, and
in the case of V$_3$Al its width is vanishing since both X and Y
atoms in the X$_2$YZ formula of the inverse Heuslers are identical
and the energy levels in the upper panel of figure 3 in Ref.\,\onlinecite{GSP}
are much closer closing the gap separating the
$t_{2g}$ and the $t_{1u}$ states. This explains the gapless
character of V$_3$Al.

\section{Conclusions}
\label{sec4}

Heusler compounds are widely studied due to their variety of
magnetic properties which they present. Especially for spintronics
and magnetoelectronics, properties like half-metallicity or
spin-gapless semiconducting behavior are crucial to enhance the
functionalities of realistic devices. This interest has been
triggered both by the advances in computational materials science,
which permit the accurate modelling of several complex materials,
and the advances in the synthesis and growth of materials which
enables the growth of materials in new metastable structures. To
this aspect the prediction of antiferromagnetic gapless behavior
in the case of V$_3$Al adopting the metastable Heusler structure
(Ref.\,\onlinecite{GalanakisSGS}) and its successful growth (Ref.
\onlinecite{Jamer}) pave the way for the incorporation of this
materials in realistic devices.

We have provided an extensive study on the electronic and magnetic
properties of V$_3$Al compound as well as its isovalent V$_3$Ga
and V$_3$In compounds using state-of-the-art electronic structure
calculations. All compounds prefer the gapless G-type antiferromagnetic
structure. The large absolute values of the spin magnetic moments of
the V atoms having antiparallel spin moments lead to a strong
short-range exchange interaction and consequently to high value
of the N\'eel temperature which approaches or even exceeds 1000 K
making them operational at room temperature. The G-type
antiferromagnetism is stable with respect to the ferromagnetic
configuration and it proves to be of strongly itinerant character.
Finally, we discussed the origin of the gapless behavior. We have
shown that the character of the bands is similar to the inverse
Heusler compounds studied in Ref. \onlinecite{GSP} but the
presence of three V atoms at the unit cell closes the gap leading
to the gapless character.

We expect our study on V$_3$Al, V$_3$Ga and V$_3$In compounds to
enhance further the interest on them. Based on our results and the
experiments in Ref. \onlinecite{Jamer}, further experiments are
needed to establish a grown mechanism optimizing the magnetic and
structural properties of these compounds in the metastable Heusler
structure leading to their eventual incorporation in devices.

\end{document}